\documentclass[referee,a4paper]{aa-5-mpa}
\usepackage{txfonts}
\usepackage{graphicx}
\usepackage{natbib}
\usepackage{amssymb}

\usepackage{natbib}
\bibpunct{(}{)}{;}{a}{}{,}  

\newcommand{\Iso}[2]{^{#1}{\rm #2}}

\newcommand{\msun}{M_\odot}

\newcommand{\logab}[2]{\mathrm{[#1/#2]}}
\newcommand{\hechr}{HE~0107--5240}
\begin{document}

\title{Models for extremely metal-poor halo stars}

\author{A.~Weiss\inst{1} \and H.~Schlattl\inst{1,2} \and
M.~Salaris\inst{2} \and S.~Cassisi\inst{3}}

\institute{Max-Planck-Institut f\"ur Astrophysik,
           Karl-Schwarzschild-Str.~1, 85748 Garching,
           Federal Republic of Germany
           \and
           Astrophysics Research Institute, Liverpool John Moores University,
           Twelve Quays House, Egerton Wharf, Birkenhead, CH41 1LD, UK
           \and 
           INAF -- Osservatorio Astronomico Collurania, 
           Via Mentore Maggini, 64100 Teramo, Italy 
           }

\offprints{A.~Weiss; (e-mail: weiss@mpa-garching.mpg.de)}
\mail{A.~Weiss}

\date{Received; accepted}


\abstract{Two alternative scenarios concerning the origin and
  evolution of extremely metal-poor halo stars are investigated. The
  first one assumes that the stars have been completely metal-free
  initially and produced observed carbon and nitrogen overabundances
  during the peculiar core helium flash typical for low-mass
  Population~III stars. The second scenario assumes that the initial
  composition resulted from a mixture of primordial material with
  ejecta from a single primordial supernovae. Both scenarios are shown
  to have problems in reproducing C, N, and O abundances
  simultaneously, and both disagree with observed
  $\Iso{12}{C}/\Iso{13}{C}$-ratios, though in different directions. We
  concentrate on the most iron-poor, carbon-rich object of this class,
  \hechr, and conclude, that the second scenario presently offers the
  more promising approach to understand these objects, in particular
  because evolutionary tracks match observations very well.
}
\maketitle

\newpage

\section{Introduction}

The extremely (or ultra) metal-poor stars (UMPS) of the galactic halo are
believed to be the closest links of the Galaxy to the first generation
of stars, to Population~III. We therefore hope to learn about the
first epoch of star formation and the end of the Dark Ages because
they either are members of 
Population~III themselves or because they carry the immediate imprint
of massive Pop.~III stars or primordial Supernovae. 
They have received considerable attention in the recent past because
of the fact that they are at the crossroads of stellar evolution, star
formation, galactic chemical evolution and cosmology, notably here the
recent CMB results and the question of reionization by Pop.~III stars.

The discovery of \hechr, with a record low abundance of heavy
elements of $\mathrm{[Fe/H]} = -5.3$, which is about a factor of 10
below the previously known lowest value, raised our interest to link
this star to Pop.~III. Remarkably, the total ``metallicity'' of \hechr\ 
is by far not metal-poor due to a carbon and nitrogen overabundance of
$\mathrm{[C/Fe]} = 4.0$ and $\mathrm{[N/Fe]} = 2.3$, putting it into a large
subgroup of UMPS with similar peculiar composition.

From the point of view of stellar evolution theory metal-free stars
are interesting in themselves because of some aspects of their
structure and evolution which differ drastically from that of ordinary
Pop.~II or I stars. One of these pecularities is the fact that during
the core helium flash mixing between the helium and hydrogen shell and
the convective envelope can take place, resulting in a carbon- and
nitrogen-rich envelope and a second red giant branch phase. Therefore,
it is a plausible assumption that the UMPS are true Pop.~III stars and
that the carbon-rich subgroup consists of stars that experienced the first,
peculiar helium flash. We want to emphasize that in this paper we are
concerned only with the abundances of the CNO-elements, since 
heavier elements are unaffected by the nuclear
processes in low-mass stars.

In our previous papers, we have therefore investigated the scenario
mentioned above to explain the observed chemical composition of
carbon-rich UMPS qualitatively and thus to link them 
to Population~III. The main problem one faces is that the
flash-induced mixing appears to result always, independent of the
details and assumptions of the calculations, in the same amount of
carbon and nitrogen, such that for $\mathrm{[Fe/H]} \lesssim -3$ the
predicted carbon overabundance is $\mathrm{[C/Fe ]} \gtrsim 3$,
which is at the upper limit of observed
values. Since the C overabundance of \hechr is as high as
the value found in our previous calculations, it appeared to be
particularly worthwhile to apply our approach to this star. Similar
calculations have recently been performed independently by
\cite{pclp:2004}.\footnote{Our results were presented at the First~Stars~II
meeting in May 2003; \\ see {\tt
http://www.astro.psu.edu/users/tabel/II/presentations/weiss.pdf}.}

We therefore present a model for \hechr, based on the flash-induced
mixing (FIM) in Sect.~2, repeating the basic features of this event. In
Sect.~3, we will then show calculations for an alternative scenario,
which assumes that this star (and other UMPS) are formed directly from
the ejecta of Pop.~III supernovae, i.e.\ that UMPS are the immediate
successors of true, massive Pop.~III objects. This scenario is also
applied to other extremely metal-poor halo stars. Sect.~4 finally
summarizes our conclusions. 

\section{A Pop.~III model for \hechr}

The stellar parameters and chemical abundances of \hechr\ (see
Table~\ref{t:1}) are taken
from \cite{ChrBB:2002}, \cite{ChrGus:2003}, and \cite{BeChrGu:2004}.
The value for oxygen is probably an
upper limit and the error given is one-sided according to
\cite{BeChrGu:2004}. The final value might be close to $\logab{O}{Fe}
= 2.0$.

\begin{table*}
\caption{Stellar parameters and chemical composition of \hechr\
  \citep{ChrGus:2003}, and two theoretical models as explained in
Sects.~2 (M1) and 3 (M2). The column ``M1 (initial)'' denotes the
composition of the polluting SN ejecta; the interior of model M1 has
$Z=0$ throughout.} 
\protect\label{t:1}
\begin{flushleft}
\begin{tabular}{l|rr|rr|rr}
\noalign{\hrule}
Qnty. & value & $\sigma$ & M1 (initial) & M1 (final) & M2 (initial) & M2 (final)\\
\noalign{\hrule}
$T_\mathrm{eff}~(K)$ & 5100 & 150 & --- & 4520 & --- & 5026\\
$\log g$ (cgs) & 2.2 & 0.3 & --- & 1.7 & --- & 2.3 \\
$M/\msun$ & $\approx 0.8$ & --- & 0.820 & 0.817 & 0.810 \\
\noalign{\hrule}
$\logab{Fe}{H}$ & -5.3 & 0.2 & -1.3 & -5.3 & -5.3 & -5.3 \\
$\logab{C}{Fe}$ & 4.0 & 0.3 & 2.6 & 6.0 & 4.0 & 4.0 \\
$\logab{N}{Fe}$ & 2.3 & 0.2 & -0.6 & 6.2 & 0.0 & 2.9 \\
$\logab{O}{Fe}$ & 2.4 & 0.4 & 2.6 & 2.7 & 4.0 & 4.0 \\
$\Iso{12}{C}/\Iso{13}{C}$ & $\approx 60$ & ($> 50$) & $\approx 10^6$ &
4.8 & $\approx 10^6$ & 61 \\
\noalign{\hrule}
\end{tabular}
\end{flushleft}
\end{table*}

\begin{figure*}[h]
\includegraphics[scale=0.7]{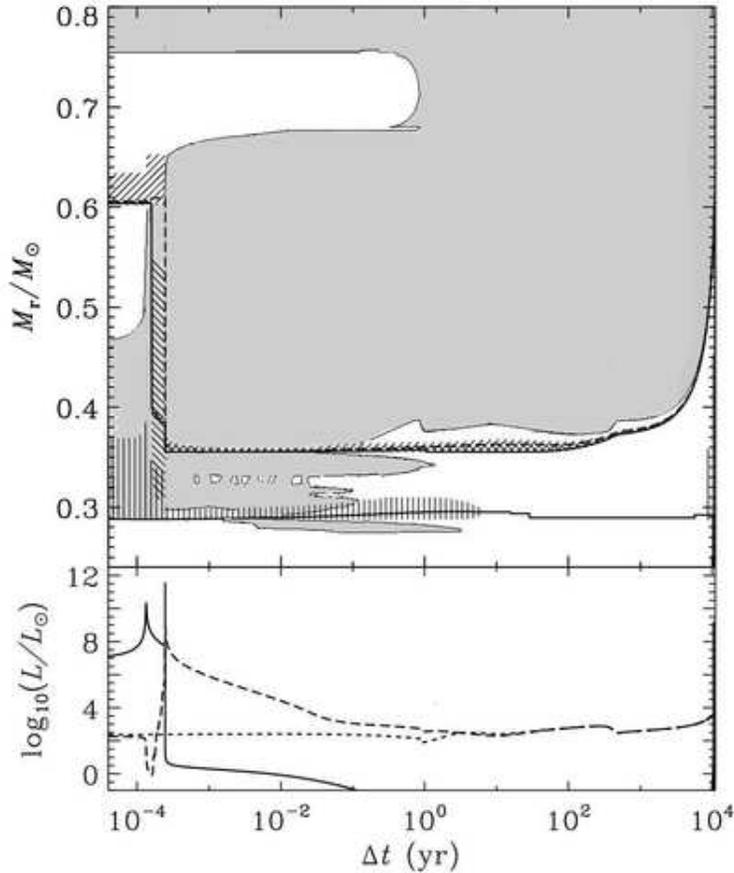}
\caption[]{Upper panel: Evolution of the nuclear shells and convective
  layers (blue, shaded regions) in an
  initially metal-free model for \hechr\ of $0.82\,\msun$ during and after the core
  helium flash, illustrating the flash-induced mixing. The (blue)
  dashed line denotes the maximum of H-burning via 
  pp-chains, the (black) solid one that via CNO-cycle. The lower (red) solid
  line is the maximum of. He-burning. Lower panel: Luminosities from
  He-burning (red, solid), H-burning (blue, dashed), and total (black,
  dotted). Time is in years since $\log L_\mathrm{He}/L_\odot \geq 6$,
  i.e.\ since the He-flash has fully started.}
\protect\label{f:1}
\end{figure*}

In our previous papers \citep{wcschs:2000,scsw:2001,sscw:2002} we
followed the idea that the UMPS are proper Pop.~III stars, i.e.\ their
initial composition was completely metal-free ($Z=0$), and that the
observed surface metal abundances, in particular that of heavy elements and
iron, are due to an external pollution or accretion event which took
place in the early phases of the approximately 12~Gyr of main-sequence
lifetime. Technically, we added a specific amount of {\em solar
  metallicity} material on top of the initial, zero-age model such
that during the Red Giant phase, after dredge-up and dilution the
observed $\logab{Fe}{H}$ values were reached
\citep[see][Paper~I]{wcschs:2000}. With this approach, the interior,
nuclear evolution of the model is that of a metal-free star.
In this scenario the peculiar C and N
abundances are then produced by the star itself during the core helium
flash. This flash-induced mixing (FIM) has been described extensively
by various authors \citep[Paper~II]{FIH90,FII00,scsw:2001}, therefore
it suffices to repeat that the convective region caused by the helium
flash is able to extend well into the hydrogen burning shell and to
mix protons into the carbon-helium intershell region, where the protons 
are immediately captured on carbon nuclei to form nitrogen. 
The additional energy from this CN-flash leads to further expansion of 
the intershell layer such that the hydrogen and helium shells are
extinguished at their previous locations and instead a rejuvenated
hydrogen shell establishes itself within the previous intershell
layer. The timescale for this event is of order of less than one day,
such that the $\Iso{15}{N}(p,\gamma)\Iso{16}{O}$ reaction is too slow
to strongly affect the oxygen abundance. Additionally, there are not
even enough protons to allow the $\Iso{12}{C}$/$\Iso{14}{N}$
equilibrium value of roughly 0.1 at this temperature to be reached.
Later on, the convective envelope is able to penetrate into the
CN-rich layers, mixing these elements to the surface. Figure~\ref{f:1}
shows the whole sequence of events for a $0.82\,\msun$ model of
$Z_i=0$ during the first year after the flash; at this time the mixing
to the surface has taken place and the nuclear shells have reached
their final structure.

\cite{FII00} already specified conditions under which the FIM could
occur: broadly speaking, a (total) metallicity of $\logab{M}{H}
\lesssim -4.5$ (where the M stands for the global metal abundance)
and mass below $1\,\msun$ are required. In
Paper~II we added the influence of such parameters as initial helium
abundance, the amount of polluting material, and the effect of
sedimentation on the main sequence. We found that the total
metallicity might be slightly higher $\logab{M}{H}  \lesssim -4$ than
quoted by \cite{FII00}, also due to the use of updated plasma neutrino
emission rates. The reader should also refer to
\cite{pclp:2004} for a detailed description of the FIM and further
tests concerning the conditions under which it can occur.

\citet[Paper~III]{sscw:2002} finally investigated in detail whether
and how the
resulting envelope composition after the FIM is influenced by a very
small, but non-zero initial metallicity, or how it depends on details
of convection theory, such as, for example, the inclusion of
overshooting. We found that final abundances of $\logab{C}{Fe}\approx 
\logab{N}{Fe} \approx 4$ and $\logab{O}{Fe} \approx 0.7$ (starting
with an $\alpha$-element, i.e.\ oxygen, enhancement of +0.4) are very
insensitive to all these variations with the exception of an artifical
reduction of the
convective mixing velocity by a significant factor of $10^{4}$  or
more. This could be the typical mixing velocity in semiconvective
layers. For example,  \cite{merry:95} found in numerical
simulations of semiconvection mixing velocities of up to
$10^3\,\mathrm{cm/s}$, while Achatz, M\"uller \& Weiss (2004, in
preparation) quote fully convective velocities of
$10^7\,\mathrm{cm/s}$ in their multidimensional hydro-simulations of
the core helium flash. Interestingly, \cite{herwig:2002} states that
in order to model ``Sakurai's Object'' (V3443~Sgr), where a similar
mixing-and-burning might have taken place, mixing velocities reduced
by a factor of $10^4$ are needed, too, in order to reproduce the
peculiar surface composition. Nevertheless, it remains to be
investigated whether the inclusion of 
semiconvection would modify our results in the desired direction. 
Similarly, a final $\Iso{12}{C}/\Iso{13}{C}$ ratio of less than 6 was
the rule, slightly lower than the quoted numbers in the literature,
which were around 10. Since hardly any 
oxygen is produced during the FIM we concluded that
$\logab{O}{Fe}$-values would be necessary to decide whether the
FIM-scenario could be the explanation for the observed anomalies. 

In the past we had assumed that the additional envelope material from
pollution is of solar composition, both for simplicity and because we
were interested rather in the interior evolution than in detailed
surface abundances. Modelling a particular star, of course, warrants a
realistic composition of the polluting material. 
The $\Iso{12}{C}/\Iso{13}{C}$ isotope ratio of \hechr\ 
\citep[$>30$;][]{ChrBB:2002} is
definitely higher than the values in our published FIM-models. In them,
it is $\approx 5$, somewhat larger than the CN-equilibrium values found in
the intershell layers. This slight enhancement is due to the
assumption that the additional surface material had solar
composition with a carbon isotope ratio of 90.
Obviously, a still higher value will also raise the final ratio of the
model, bringing it in better agreement with observations.

\cite{umno:2003} showed that the pattern of elements heavier than Mg in
\hechr\ agrees
with that predicted from nucleosynthesis in zero-metallicity type II
supernovae of initial mass 
$20-130\,\msun$ \citep{umno:2002}, under the assumption of severe fall-back to a
massive black hole and complete mixing of the ejected material. 
Specifically they use a $25\,\msun$ model with explosion energy of
$3\cdot 10^{50}\,\mathrm{erg}$, complete mixing of material within the
helium core of $6\,\msun$ and severe fallback of matter interior to
$1.8\,\msun$, such that only $8\cdot 10^{-6}\,\msun$ of $\Iso{56}{Ni}$
(or Fe, after the nuclear decay) are ejected. The latter condition
comes from the requirement that $\logab{C}{Fe} \approx 4$ is
achieved. 

We therefore took the composition for the polluting material to be that of
a theoretical Pop.~III supernova explosion from \cite{chl:2002},
selecting the $35\,\msun$ model. The explosion energy in this model
had been set to $1.2\cdot 10^{51}\,\mathrm{erg}$. The results of
\cite{chl:2002} do not 
differ significantly from those of \cite{umno:2002} for comparable
parameters (progenitor mass, mass cut, explosion energy), at least with respect
to the elements of interest for us. In this SN model, a total of
$32.5\,\msun$ is ejected, of which there are $14.7\,\msun$ of
hydrogen, $10.7\,\msun$ of helium, $1.2\,\msun$ of carbon,
$32.2\cdot10^{-4}\,\msun$ of nitrogen, and $3.6\,\msun$ of oxygen. This matter
is assumed to be completely mixed.
The amount of polluting material actually dumped onto the initial model
and the mass cut have to be selected such that
the final iron abundance on the RGB after dredge-up and helium flash
mixing matches that of \hechr. Note that we do not have to care about
$\logab{C}{Fe}$ in the polluting material, since carbon production
during the helium flash and subsequent mixing will dominate the final
abundances. In model ``M1'' presented here (see
Table~\ref{t:1} and Fig.~\ref{f:1}), the $\Iso{56}{Ni}$ mass was
$0.001\,\msun$, and the amount 
of polluting material $6\cdot 10^{-5}\,\msun$. Its 
initial composition is given in
Table~\ref{t:1}, column ``M1 (initial)''.
While $\mathrm{[Fe/H]}
= -1.3$ in the SN ejecta, this is reduced to $-3.6$ already in the
initial main sequence model, which has a convective envelope of
$0.016\,\msun$. After the convective envelope has reached is deepest
extent of about $0.4\,\msun$, the correct final $\logab{Fe}{H}$ of
-5.3 is reached. We have also
run models with a $\Iso{56}{Ni}$ mass increased respectively reduced
by a factor of 10, implying opposite factors for the additional
polluting mass. 

The core helium flash and the induced mixing that takes place are
illustrated in Fig.~\ref{f:1}. The resulting abundances after the
flash can be found in Table~\ref{t:1}, column ``M1
(final)''. Obviously, there is again too much C and N produced, and
$\Iso{12}{C}/\Iso{13}{C}$ is much too close to the equilibrium value,
because of the small amount of polluting matter added to the star. As
stated before, we have modified this parameter, but found only
marginal variations in the resulting abundances, except in the most
extreme case of $6\cdot 10^{-3}\,\msun$ of polluting material, with
only $10^{-4}\,\msun$ of $\Iso{56}{Ni}$. In this case, the final C and
N abundances were lower by one of order of magnitude due to the large
amount of polluting matter and the carbon isotope ratio increased
slightly to 5.0. However, in this case $\logab{Fe}{H} = -4.2$ is too
high. While one could try to find better suited initial SN-yields, for
example by choosing another SN-progenitor mass, inspection of the
corresponding tables in the papers quoted reveal that in order to
achieve the very low Fe abundance one always has to add such small
amounts of SN-ejecta that the dilution of the carbon-rich intershell
matter of approximately $0.3\,\msun$ by the carbon-poorer envelope is
almost negligible, such that the final C and N abundances can hardly
be reduced 
to the level of \hechr. Additionally, the $\Iso{12}{C}/\Iso{13}{C}$ is
always too low, even if the SNe is basically $\Iso{13}{C}$-free. We
therefore conclude that -- as in the case of other, more iron-rich
UMPS -- the amount of carbon and nitrogen relative to iron is too
high. In fact it appears that in the observed stars $\logab{C}{H}$ and
$\logab{N}{H}$ are constant within a factor of 10, and that this value
is lower by a factor of 10---100 than that of the models. In addition,
the carbon isotope ratio reflects that the observed material has been
exposed to CN-burning to a much lower degree than our models
predict. The oxygen abundance in M1, which nicely fits that of \hechr\ is
solely due to the initial SN-composition, since hardly any oxygen is
produced in the model.

\section{Pop.~II.5 models}

\subsection{Model for \hechr}

\cite{umno:2003} and \cite{numodm:2003} advocate the idea, based on
the heavy element abundances, that \hechr\ and other UMPS are second
generation stars, forming immediately after only one Pop.~III
supernova has exploded and thus carrying the immediate imprint of
it. Such objects have also been termed ``Pop.~II.5'', to discriminate
them from Pop.~II, where the heavy metal composition is the result of
many, well mixed SNe.  We add that \cite{lcb:2003} critized
this idea on grounds of incompatible Ni and C abundances, which
require very different mass cuts and because of the relative
abundances among lighter elements like Na and Mg. Instead, they suggest
the superposition of two primordial supernovae of 15 and
$25\,\msun$. Since we are concerned here with only a few elements (C,
N, O, Fe), this alternative scenario, which predicts a too large
oxygen abundance of $\logab{O}{Fe} = 4.1$, does not differ
qualitatively for our purposes, such that we follow the simpler
suggestion of \cite{umno:2003}.  Both scenarios imply that \hechr\ had
a homogeneous initial composition with a heavy metal abundance
throughout the interior as is observed today, in contrast to the model
of the last section, which had this only in the polluted envelope
layers. We therefore calculated the straightforward evolution of such
a model (M2) up the RGB and through the helium core flash.

For the SN model we chose the same $35\,\msun$ model by
\cite{chl:2002} as before. In order to obtain the observed
$\logab{C}{Fe}$, a mass cut of $M(\Iso{56}{Ni}) = 6\cdot 10^{-5}\,\msun$ is 
necessary, and a dilution of 1:360 with primordial H/He-matter to
arrive at $\logab{Fe}{H} = -5.3$. All other element abundances are
resulting from the SN model (see Table~\ref{t:1}). In particular, the
initial nitrogen abundance is approximately  solar, and
$\Iso{12}{C}/\Iso{13}{C}$ very high.

\begin{figure*}
\includegraphics[draft=false,scale=0.45]{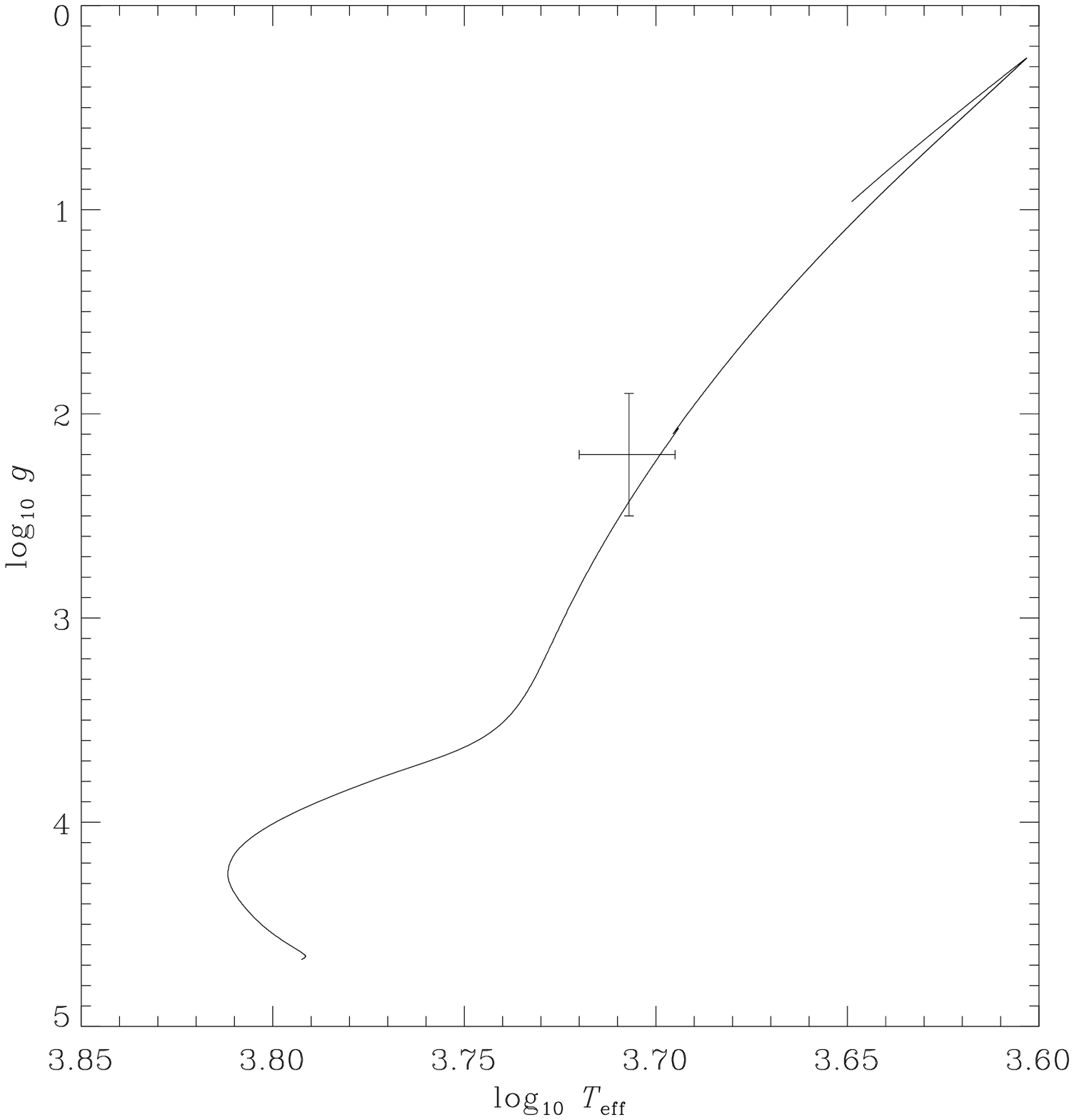}
\caption[]{Evolution of model M2 ($\log g$ vs.\ $\log T_\mathrm{eff}$), 
which is a star of $0.81\,\msun$ and
initial iron abundance of $\logab{Fe}{H}=-5.3$ (see text for
details). The error cross marks the position of \hechr.}
\protect\label{f:2}
\end{figure*}

The main sequence and red giant evolution is that of an ordinary, 
moderately metal-poor low-mass star. The mass was chosen as
$0.81\,\msun$ to obtain an age of 13.1~Gyr at the RGB tip.
During core hydrogen burning
CN-conversion takes place and the nuclear results become evident after
the first dredge-up. The final nitrogen abundance as well as the
carbon isotope ratio are quite similar to those observed in
\hechr\ (Table~\ref{t:1}). Oxygen is hardly changed during the
evolution, and reflects completely the choice of the SN
model. A survey of the quoted Pop.~III SNe model literature reveals
that $\logab{O}{C}$ varies between -0.2 and 1.1; in our selected model
it is close to the solar value. The very high carbon
abundance of \hechr\ thus implies a similar oxygen enrichment, which is
therefore too abundant by more than one order of
magnitude and poses one of the problems of the present scenario for
the nature of \hechr. 
Figure~\ref{f:2} shows the
evolutionary track of M2 (in the $\log g$ vs.\ $\log T_\mathrm{eff}$
diagram), on top of which the observed position of \hechr\ and error
bars \citep{ChrGus:2003} are plotted. Obviously, there is a very good
agreement, and 
indeed \hechr\ is in a post-dredge-up phase. The assumption that
\hechr\ is a star on the first red giant branch, which has evolved
as a single star, is supported by the result of \cite{BeChrGu:2004}.

Due to the rather high total metallicity, which is one of the basic
parameter influencing the appearance of the FIM event, the core helium
flash happens at high luminosity and without any non-canonical or extended
mixing. 

We close this part with a comment on opacities. So far we have used
tables with appropriate H, He, and metal abundances, where the
internal metal distribution is assumed to be that of $\alpha$-element
rich Pop.~II stars. Indeed, in the present models a large part of the
metals is actually carbon and nitrogen. Since the total metallicity is
$Z=7.1\cdot 10^{-4} = 0.037\, Z_\odot$, one can no longer assume that
the individual metal abundances are unimportant \citep[see][for a
discussion of this issue]{sw:98}. As in \cite{scsw:2001} we have
therefore also computed a model with opacities for C- and N-enhanced
matter (see this paper for detail), and found no significant
differences, except that the model is younger by 0.7~Gyr after the main
sequence. Due to lack of appropriate tables we could not account for
the increased oxygen abundance.

\subsection{Models for other objects}

In addition to \hechr\ we calculated additional models for several
other objects with abundances taken from the literature (see
Table~\ref{t:2}), which are believed to be first ascent giants on
grounds of their surface gravity and effective temperature. The
composition of the initial, homogeneous model 
was always obtained by mixing ejected material of the $35\,\msun$
supernova model by \cite{chl:2002} with pristine matter. The mass cut
and mixing factor are the free parameters chosen in such a way as to
obtain approximately the observed iron and carbon
abundance. 

\begin{table*}[h]
\caption{Observational data for our selection of extremely metal-poor
  halo stars; \hechr\ is repeated for completeness} \label{t:2}
\begin{tabular}{c|ccccccc|l} \hline \hline
object & ${\rm [Fe/H]}$ & $T_{\rm eff}$ & $\log g$ & ${\rm [C/Fe]}$ & ${\rm
[N/Fe]}$ & ${\rm [O/Fe]}$ & ${\rm ^{12}C/^{13}C}$ & reference \\ \hline
HE~0107-5240 & $-$5.30 & 5100 & 2.2 & 4.0 & 2.3 & 2.4 & 60 &
\cite{ChrGus:2003} \\
& & & & & & & & \cite{BeChrGu:2004} \\ \hline
CS~22943-037 & $-$4.00 & 4900 & 1.5 & 1.2 & 2.7 & 2.0 & 4 & \cite{dhssp:2002}\\
CS~29498-043 & $-$3.75 & 4400 & 0.6 & 1.9 & 2.3 & --- & 6 & \cite{anrba:2002}\\
CS~22957-027 & $-$3.11 & 5100 & 1.9 & 2.4 & 1.6 & --- & 8 & \cite{anrba:2002}\\
CS~22892-052 & $-$2.97 & 4850 & 1.5 & 1.1 & 1.0 & 0.7 & --- &
\cite{nrb:97} \\
& & & & & & & & \cite{sclibb:2003} \\
CS~31082-001 & $-$2.90 & 4850 & 1.5 & 0.2 & $<0.2$ & 0.6 & $<20$ & \cite{hpcbn:2002}\\
\hline 
\end{tabular}
\end{table*}

\begin{figure*}[h]
\includegraphics[draft=false,scale=0.45]{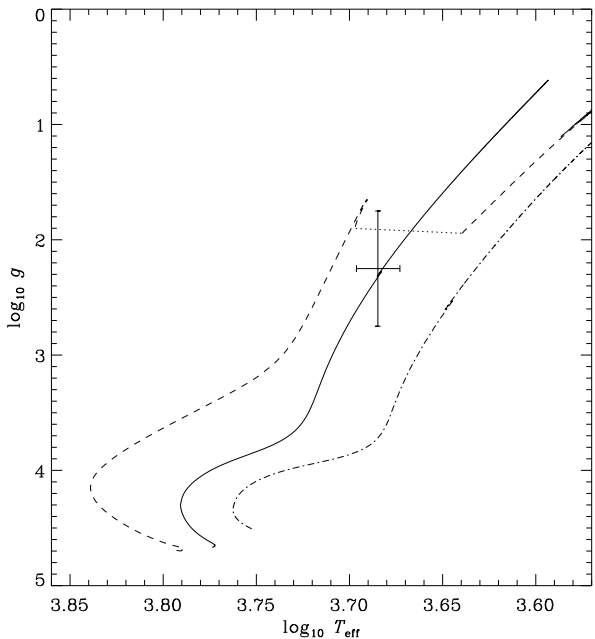}
\caption[]{Evolution of our model for CS22957-027 (Table~\ref{t:3};
solid line)
  compared to that of a Pop.~III model similar to M1 ($0.82\,\msun$;
dashed)
experiencing the flash-induced mixing at the tip of the first RGB  and
one   of a $1\,\msun$ star of typical Pop~I composition (dotted line).}
\protect\label{f:3}
\end{figure*}

Figure~\ref{f:3} shows the resulting evolutionary track of the model
for CS22957-027 together with the Pop.~III model of Sect.~2 (dashed
line) and a
solar-type Pop.~I star to emphasize the effect of the strong C and N
enrichment after the FIM. Since $T_\mathrm{eff}$ and $\log g$ of
\hechr\ are very similar to that of CS22957-027, this plot also
illustrates the fact that the post-flash Pop.~III model is slightly
too cool for \hechr.

Table~\ref{t:3} contains the summary of these
calculations. Since the evolutionary tracks cross the observationally
allowed range in $\log g$ and $T_{\rm eff}$ several times, we
offer various possible choices along the RGB and AGB evolution. We
note in particular that some stars have high enough surface gravities
to be still on the lower RGB, before the hydrogen shell encounters the
composition discontinuity left behind by the convective envelope when
it had reached its deepest extension. This phase is commonly refered
to as the ``bump'', because of the slower evolutionary speed of red
giants when they reach the suddenly increasing hydrogen supply above
the discontinuity \citep[see][for a review on low-mass red giant
evolution]{scw:2002}.

Comparing the final abundances with those observed it becomes obvious
that no star can easily be modeled. For all of them the carbon isotope
ratios are much too high; we are therefore facing the opposite problem
we have with the FIM-scenario. In addition, nitrogen abundances for
most stars, but in particular for
CS~22943-037 and CS~31082-001 are too low, indicating a lack of
mixing. Both problems could be cured if one assumes that these stars
experience at or after the bump additional mixing between the hydrogen
shell and the convective envelope. Such extra-mixing is now known to
take place in basically all low-mass metal-poor stars, both in the
field and in clusters, and leads to reduced carbon and increased
nitrogen abundances as well as to reduced carbon isotope ratios. We
refer the reader to \cite{gsc:00} for an observational overview and to
\cite{ddn:98} for a comprehensive theoretical paper.\footnote{A
summary of state-of-the-art-data and ideas will be found in the
proceedings of Joint Discussion 4 (Astrophysical impact of abundances
in globular cluster stars) of the IAU General Assembly XXV
(Sydney, 2003), edited by D'Antona et al.}

The bottom two lines of Table~\ref{t:3} refer to a case, for which we
used a $15\,\msun$ model by \cite{umno:2002} instead of our standard
SN-model, to show the influence of the initial composition. The iron
abundance of these ejecta is too high ($\logab{Fe}{H} = 0.7$;
$\logab{C}{Fe}=0.0$)  for the required
$\logab{Fe}{C}$, such that we multiplied it by a factor 0.067 (given
in column 4). Nitrogen and oxygen have abundances of -0.6 and -0.1 in
the standard spectroscopic scale. The explosion energy was
$10^{51}\,\mathrm{erg}$. In spite of the different initial composition
there is hardly a change in the final one, with the exception of the
nitrogen abundance on the RGB.

\begin{table*}
\caption{Pop.~II.5 models using the $M = 35 M_\odot$ SN yields of
\cite{chl:2002}. The mass cut has been found by inter- or extrapolating
between SN models with different ${M_{\rm SN}({\rm Ni})}/{M_\odot}$ 
to obtain the correct ${\rm [C/Fe]}$. ${M({\rm prim})}/{M({\rm
SN})}$ is the mixing proportion between primordial and SN matter,
needed to reach (approximately) the observed ${\rm [Fe/H]}$-value.
The various parameters of the models start at column 5 (age).
Evolutionary models that could represent the
objects are marked by the following labels:
E-RGB: on RGB, but before bump; 
B-RGB: at bump; AB-RGB: after bump; L-RGB: close to tip; E-AGB: on
early AGB, after 2$^{\rm nd}$ dredge-up; L-AGB: immediately before
thermal pulses commence; TP-AGB: thermal pulse phase on AGB.
The last two lines refer to models for CS~22943-037, for which we have
used SN-yields from a $15\,\msun$ model by \citealt{umno:2002} (see
text).} \label{t:3} 
{\tiny
\begin{tabular}{cccccccccccccr} \hline \hline
object & $\frac{M_\star}{M_\odot}$ & $\frac{M({\rm prim})}{M({\rm SN})}$ & 
$\frac{M_{\rm SN}({\rm Ni})}{M_\odot}$ & $\frac{\rm Age}{\rm Gyr}$ &
  ${\rm [Fe/H]}$ &  $T_{\rm eff}$~[K]
& $\log g$ & $\log\frac{L}{L_\odot}$ & {\small ${\rm [C/Fe]}$} & {\small${\rm
[N/Fe]}$} & {\small ${\rm [O/Fe]}$} & $\frac{\rm ^{12}C}{\rm ^{13}C}$
\\ \hline
{HE~0107-5240} & 0.81 & 350 & $6\times10^{-5}$ & 13.1 & $-$5.31 & 5026 &
2.3 & 1.8 & 4.0 & 2.9 & 4.0 & 61 & {\small E-RGB} \\[1mm]
{CS~22943-037} & 0.79 & 10500 & 0.06 & 13.5 & $-$4.03 & 4880 &
1.5 & 2.5 & 1.2
& $-$1.8 & 1.2 & 340 & {\small AB-RGB} \\ 
& & & & 13.6 & $-$4.02 & 4800 & 1.2 & 2.8 & 1.1 & 1.1 &
1.2 & 56 & \small{E-AGB} \\[1mm]
{CS~29498-043} & 0.79 & 1170 & 0.1 & 13.7 & $-$3.77 & 4380 &
0.7 & 3.2 & 1.9 & 0.1 & 1.9 & 97 & {\small L-RGB}\\ 
& & & & 13.8 & $-$3.77 & 4400 & 0.6 & 3.2 & 1.8 & 1.9 &
1.9 & 43 & {\small L-AGB} \\
& & & & 13.8 & $-$3.77 & 4400 & 0.6 & 3.2 & 1.8 & 2.0 & 1.9 &
43 & {\small TP-AGB}\\[1mm]
{CS~22957-027} & 0.85 & 84 & 0.003 & 13.5 & $-$3.12 & 4800 &
2.4 & 1.7 & 2.4 & 1.7 & 2.4 & 55 & {\small B-RGB}\\[1mm]
{CS 22892-052} & 0.795 & 1230 & 0.075 & 13.4 & $-$2.99 & 4790 & 1.5 &
2.5 & 1.1 & $-$0.6 & 1.1 & 94 & {\small AB-RGB} \\ 
& & & & 13.5 & $-$2.99 & 4740 & 1.1 & 2.8 & 1.0 & 0.9 & 1.1
& 33 & {\small E-AGB}\\[1mm]
{CS~31082-001} & 0.79 & 0.79 & $0.6$ & 13.5 & $-$2.93 & 4850 &
1.5 & 2.5 & 0.2 & $-$2.7 & 0.2 & 280 & {\small AB-RGB} \\ 
\hline 
{CS~22943-037} & 0.79 & 2720 & 0.067 & 13.5 & $-$4.00 & 4920 & 1.5 & 2.5
& 1.2 & 0.6 & 1.1 & 370 & {\small AB-RGB} \\
& & & & 13.6 & $-$4.00 & 4833 & 1.3 & 2.8 & 1.1 & 1.1
& 1.1 & 59 & {\small E-AGB}\\
\hline
\end{tabular}
}
\end{table*}

\section{Discussion}

In earlier papers (Papers II and III) we have applied our scenario for
the evolution of initially 
metal-free Pop.~III stars with additional surface pollution to some
known objects of the galactic halo, investigating the possibility that
the observed severe carbon and nitrogen enhancements are due to
internal production and mixing in the course of the first core helium
flash. These models indicated that both the abudnances of C and N in
the models are too high, and that the carbon isotope ratio is too
close to equilibrium values. In addition, we noticed at that time that
a statistically representative sample of UMPS is needed to verify that
the number of carbon-enhanced objects is in agreement with the
predictions from the models, which would be post-flash, and therefore
short-lived compared to lower RGB stars. The detection of \hechr\ with
the lowest iron abundance of $\logab{Fe}{H} = -5.3$ and highest carbon
enhancement of $\logab{C/Fe} = 4.0$, allowed us to investigate this
scenario more closely. We thus computed specific models using
realistic SN yields from the literature.

The low iron content of \hechr\ can only be achieved by adding tiny
amounts of SN-ejecta matter or to impose rigid mass cuts for the SN
explosion. We find that, independent of the particular choice of
polluting matter, the final carbon and nitrogen abundances, which
result from the core helium flash and the subsequent
mixing, exceed the observed abundances by orders of magnitude,
similar to the case of the less extreme cases we modeled in our previous
papers. It appears that both in nature, and in our models, the {\em
amount} of overabundant C and N is rather constant, such that with
decreasing Fe-abundance the relative overabundance is increasing. This
effect can actually been seen in Fig.~2 of \cite{rbs:99} by looking at
the upper envelope of the $\logab{C}{Fe}$ vs.\ $\logab{Fe}{H}$
distribution. However, the models appear to produce 10--100 times too
much carbon and nitrogen. 
Additionally, the carbon isotope ratio in \hechr\ is definitely
far above equilibrium values, which are always obtained in our
simulations. These results, with the exception of the oxygen
abundance, are widely independent on whether we use solar or early SNe
material for the pollution. 
Therefore, \hechr\ contradicts most strongly our
flash-induced mixing models. 

We then investigated an alternative scenario, assuming that \hechr\
and other extremely metal-poor stars are representing a kind of ``early
Pop.~II'' or Pop.~II.5 class of objects. Their homogeneous initial
composition is of low, but finite metallicity. 
However, contrary to standard Pop.~II stars, the material
still carries the imprint of one or few individual SNe of Pop.~III. We
restricted ourselves to one particular SN model, since in terms of
CNO-elements the various models available (mass, explosion energy,
author) do not vary drastically. The choice was made according to
inferences based on reproducing the heavy elements in \hechr.

We find that, after mass cut and dilution with pristine interstellar
material have been fixed, the stars are carbon and oxygen rich already
on the main sequence, produce large amounts of nitrogen in CN-cycling,
and expose them as a consequence of standard first dredge-up. The
evolution is quite standard, as for Pop.~II models. C and N abundances
agree naturally very well with the observations, but carbon isotope
ratios are in this scenario definitely {\em higher} than for most stars
under consideration (with the exception of \hechr). Therefore, for 
$\Iso{12}{C}/\Iso{13}{C}$ we face the opposite problem of the one
we have for the 
FIM-models. Oxygen is always, due to the composition of
SN ejecta, enriched. It would therefore be
necessary to obtain results for oxygen for UMP stars\footnote{It
  appears that indeed several of them have $\logab{O}{Fe} \approx 2$,
  similar to \hechr\ (V.~Hill, private communication).}
to put further constraints on the various possibilities for the nature
of the UMPS. 
In case of the objects we tried to model and for which we had oxygen
abundances, it appears that we can roughly reproduce the
observation. However, in case of \hechr\ the SN yields predict too
much oxygen. This problem can also be noticed in the models by
\cite{lcb:2003}, while it seems to be less severe in \cite{umno:2002}
using a less massive SN progenitor with only $0.3\cdot
10^{51}\,\mathrm{erg}$ explosion energy. The composition of one of the
objects we investigated could be matched satisfactorily.

The observed stars lie very nicely on the tracks for our
Pop.~II.5 scenario, although the errors are too large to allow
excluding the FIM-possibility on that ground. It is possible to
identify different evolutionary stages for the observed
objects. Generally, the later the evolutionary phase, the higher the N
abundance and the lower the $\Iso{12}{C}/\Iso{13}{C}$ ratio, reducing
some of the problems. If the star would be on the AGB,
a better agreement is possible, in particular, if the 3rd dredge-up
happens during the thermal pulses, because in this case, the carbon
isotope ratio would be reduced strongly. However, we have not
followed the evolution of the models this far. 
Again, statistically significant samples of UMPs would be
necessary to further look into this question.

In spite of the remaining problems, we presently favor the idea that the observed
extremely metal-poor stars of the galactic halo are stars formed
directly from the ejecta of one or few Pop.~III SNe of intermediate mass
($15-60\,\msun$), which are diluted with metal-free primordial
gas. Overall, the agreement between model and observations appears to
be better, and there is still a large uncertainty concerning the
SN yield composition. Also, whether one or two or a few SNe have
contributed to the initial composition of an UMP \citep[see the
discussion in][]{lcb:2003}, allows further fine-tuning of
models. Nevertheless, solid statistical samples are clearly needed for
further progress. Finally, we point out that all SN models favored,
indicate progenitor masses of $\approx 20 - 60\,\msun$. Currently, the
primordial star formation scenario is favouring much higher initial
masses for Pop~III stars. This question, too, remains to be
cleared. The extremely metal-poor stars with their particular
composition, may guide us in this.

\begin{acknowledgements}
A.W.\ wishes to thank the participants of the ``First Stars II''
meeting held at Pennsylvania State University in June 2003 for
encouraging him to publish these results presented then, and the
organizers and the DFG for substantial travel support.
S.C.\  warmly thanks for financial support by MURST (PRIN2002,
PRIN2003). 

\end{acknowledgements}
\newcommand{\singlet}[1]{#1}

\end{document}